\documentclass[12pt,oneside,final]{amsart}

\usepackage[utf8x]{inputenc}

\usepackage{mystyle}

\begin{document}

\title{EMT, Pseudo-EMT and all that\\
       Improvement and Superpotentials}
\author{Klaus Sibold}
\address{Institute for Theoretical Physics, Leipzig University, Germany.\emph
{Email address: }\rm \texttt{sibold(at)uni-leipzig.de}.}

\date{\today}

\begin{abstract}
Within Enstein-Hilbert gravity, higher derivatives  and a scalar field as 
representative of matter different versions of tensorlike quantities are discussed.The concepts of improvement and superpotential help to understand the details
of their construction and meaning. On this basis it is claimed that the higher
derivatives which are necessary for defining higher orders in perturbation
theory do not ruinthe physical content of the model.\\
\end{abstract}

\maketitle

\tableofcontents

\section{Introduction}
Within Einstein-Hilbert gravity (EH) amended by higher derivative terms and 
a massless scalar field $\varphi$ we discuss 
(pseudo-)energy-momentum operators (pEMT's).  For the EH-gravity part they
have been studied by Einstein, Klein, Weyl \cite{HW}, Freud \cite{PF}, 
Goldberg \cite{JG} and Trautman \cite{AT}.
We further consider higher derivative terms, since upon quantization and 
construction of higher orders of perturbation theory they are necessary. A 
scalar field is taken into account for completeness and comparison reasons.\\
The aim is to translate the classical results into the framework of quantum field
theory. For the purposes of this paper it is sufficient to discuss vertex 
functions in tree approximation. This permits to handle all quantities as if 
they were classical.
Their quantum nature would appear only after Legendre transformation to the 
connected Green's functions.\\
Our source leading into this area of research is a survey provided by 
C.\ Denson Hill and Pawel Nurowski \cite{HN}.\\
In previous papers the present author and Steffen Pottel,
\cite{pottel2021perturbative}, \cite{PS23}, \cite{KS24}  
have discussed the
renormalization of the respective models to all orders of perturbation theory
and presented the required 
technical tools like renormalization group equation, Lowenstein-Zimmermann
equation and rigid Weyl invariance. The main obstacle when employing higher
derivative terms is the fact that those introduce negative norm in  
state space. This problem has been reduced in \cite{PS23} to tree level 
amplitudes, 
since in higher orders, beginning with one-loop the poles of these massive 
would-be particles can not be reached. In the present paper it will be proposed to
study energy-momentum operators and their relation to so-called superpotentials.
It will be seen that this sheds light on possible gravitational wave radiation.\\

\section{PseudoEMT's and EMT's}

\subsection{The matter sector}
From  \cite{EKKSIII} we recall the results for the model of one massless 
scalar field with action
\be\label{mttct}
\Gamma({\rm matter})=\int(\frac{1}{2}\partial\varphi\partial\varphi 
		   -\frac{\lambda}{4!}\varphi^4)\equiv\int\mathcal{L}_m
\ee
This action is invariant under rigid translations
\be\label{rtrs}
W^T\Gamma\equiv\int\delta^T\varphi\frac{\delta\Gamma}{\delta\varphi}=0
\qquad\delta^T\varphi=a^\mu\partial_\mu\varphi \qquad a^\mu= {\rm const.}
\ee
Rendering the infinitesimal parameter $x$-dependent $a^\mu=a^\mu(x)$
and differentiating with respect to it, we obtain the local Ward identity (WI)
which expresses the conservation of the energy-momentum tensor
\be\label{lce}
\tilde{w}^T_\mu\Gamma\equiv
\partial_\mu\varphi\frac{\delta\Gamma}{\delta\varphi}=-\partial_\nu T_\mu^\nu
\ee
It is the so-called canonical EMT
\begin{align}\label{mcemt}
T_\mu^\nu=& -\delta_\mu^\nu \mathcal{L}_m+\partial_\mu\varphi\partial^\nu\varphi\\
T_\mu^\nu=&\partial_\mu\varphi\partial^\nu\varphi
           -\frac{1}{2}\delta_\mu^\nu\partial\varphi\partial\varphi
	   +\frac{\lambda}{4!}\delta_\mu^\nu\varphi^4
\end{align}
It is conserved on-shell, i.e.\ after using the equation of motion. 
In the next step we add a total derivative contact term (tdct)
\begin{align}\label{sctdcts}
w^T_\mu\Gamma\equiv& \partial_\mu\varphi\frac{\delta\Gamma}{\delta\varphi}
-\frac{1}{4}\partial_\mu(\varphi\frac{\delta\Gamma}{\delta\varphi})
	=-\partial_\nu T^\nu_\mu\\
T^\nu_\mu=&\partial_\mu\varphi\partial^\nu\varphi
           -\frac{1}{2}\delta^\nu_\mu\partial\varphi\partial\varphi
	   -\frac{1}{4}\delta^\nu_\mu\varphi\Box\varphi\label{tfm}
\end{align}
This tensor is also conserved on-shell.
The main reason for doing so is the fact that these contact terms form an 
interesting algebra \footnote{An other one is that one can form $x$-moments
leading to the conformal algebra.}.
\begin{align}\label{lgbrct}
	W(\varphi;a)\equiv& \int a^\mu(x)w^T_\mu\\
[W(\varphi;a),W(\varphi;b)]=&W(\varphi;b^\nu\partial_\nu a-a^\nu\partial_\nu b)
\end{align}
It is non-abelian and, in fact, just the algebra for BRST transformations,
once one interprets the $a^\mu$ as the anti-commuting Faddeev-Popov fields.
I.e.\ these local translations lead immediately to the transformation law
of gravity. The relevant details will be presented below in subsection 2.3.\\

Another feature which plays a decisive role in what follows is the possibility
of ``improving'' these EMT's: one can add to them a term
\begin{align}\label{mprvt}
I^\nu\,_\mu=&c(\delta^\nu\,_\mu\Box-\partial^\nu\partial_\mu)\varphi^2\\  
	=&2c(\partial_\mu\varphi\partial^\nu\varphi
	-\varphi\partial_\mu\partial^\nu\varphi
	+\delta_\mu^\nu(\partial\varphi\partial\varphi+\varphi\Box\varphi)
\end{align}
Obviously it vanishes identically upon taking a derivative, say $\partial_\nu$.
But this also means that it can never be created via Ward identity which
expresses conservation of the respective EMT. It is to be noted that this change 
does not influence the charge associated with the current.\\
What does it improve? When added with the value $c=1/6$ to (\ref{tfm}) the trace
of the sum vanishes. It thus changes the transformation law of the current under
special conformal transformations. This sum transforms
as a conformal field with dimension four and spin two. Hence demanding this
transformation law and the improvement one has defined a unique EMT.\\

\subsection{The gravity sector}
Within the context of classical general relativity Freud, Goldberg and Trautman 
\footnote{Conventions: Weyl, Freud, Goldberg, Trautman - all use signature
(+ for $t$, - for $x$. Freud, Goldberg use label $4$ for $t$; 1,2,3 for $x$. 
Christoffel: Weyl agrees with our (= Landau-Lifschitz, 1971), Freud also. Goldberg, Trautman do notuse it in the papers we refer to.}
define in a completely naive way a canonical EMT for EH. In terms of the 
vertex functional in tree approximation this reads
\be\label{cnmt}
\partial_\mu g^{\mu'\nu'}\frac{\delta}{\delta g^{\mu’\nu’}}\Gamma_{\rm EH}=
\partial_\lambda(\sqrt{-g} t_\mu^\nu)
\ee 
An explicit expression for $\sqrt{-g}t_\mu^\nu$ has been given by Goldberg
\cite{JG}
in terms of $\tilde{g}\equiv \sqrt{-g}g^{\mu\nu}$. It is based on separating 
from EH a total divergence
\begin{align}\label{spltR}
\sqrt{-g}R=&
\partial_\kappa(\tilde{g}^{\lambda\kappa}\tilde{g}_{\rho\sigma}\partial_\lambda
\tilde{g}^{\rho\sigma}
+2\partial_\lambda\tilde{g}^{\lambda\kappa}) +\mathcal{L}\\
\mathcal{L}\equiv& g^{\mu\nu}(\Gamma^\rho_{\mu\nu}\Gamma^\lambda_{\rho\lambda}
	-\Gamma^\lambda_{\mu\rho}\Gamma^\rho_{\nu\lambda})\\
	=&\frac{1}{8}\lbrace 
	2\tilde{g}^{\rho\sigma}\tilde{g}_{\alpha\lambda}\tilde{g}_{\kappa\tau}
	-\tilde{g}^{\rho\sigma}\tilde{g}_{\lambda\kappa}\tilde{g}_{\alpha\tau}	
	-4\delta^\sigma_\kappa\delta^\rho_\alpha\tilde{g}_{\lambda\tau}\rbrace
\partial_\rho\tilde{g}^{\lambda\kappa}\partial_\sigma\tilde{g}^{\alpha\tau}
\end{align}
Note: $\mathcal{L}$ is not $s$-covariant. In the literature it is commonly
also denoted by $G$. \\
In fact, this reformulation of $\sqrt{-g}R$ admits to define
$t^\nu_\mu \,$ as the canonical EMT for gravity with $\mathcal{L}$ as 
Lagrangian, because the latter depends now only on first order derivatives of
$g^{\mu\nu}$. However it will not transform as a
tensor under $s$, but at most under linear coordinate transformations, as
the derivation from the expression of $\mathcal{L}$ in terms of Christoffel
symbols indicates.
\begin{align}\label{cMt}
	t_\mu^\nu=&-\delta_\mu^\nu\mathcal{L}
+\frac{\partial\mathcal{L}}{\partial(\partial_\nu \phi)}\partial_\mu \phi\\
t^\nu_\mu=&-\frac{1}{8}\delta^\nu_\mu\lbrace 
	2\tilde{g}^{\rho\sigma}\tilde{g}_{\alpha\lambda}\tilde{g}_{\kappa\tau}
	-\tilde{g}^{\rho\sigma}\tilde{g}_{\lambda\kappa}\tilde{g}_{\alpha\tau}	
	-4\delta^\sigma_\kappa\delta^\rho_\alpha\tilde{g}_{\lambda\tau}\rbrace
\partial_\rho\tilde{g}^{\lambda\kappa}\partial_\sigma\tilde{g}^{\alpha\tau}\\
	&+\frac{1}{4}\lbrace 2\tilde{g}^{\nu\sigma}\tilde{g}_{\alpha\lambda}
	                                      \tilde{g}_{\kappa}{\tau}
	-\tilde{g}^{\nu\sigma}-\tilde{g}_{\lambda\kappa}\tilde{g}_{\alpha\tau}
	-4\delta^\nu_\alpha\delta^\sigma_\kappa\tilde{g}_{\lambda\tau}\rbrace
	\partial_\sigma\tilde{g}^{\alpha\tau}\partial_\mu \tilde{g}^{\lambda\kappa}
\end{align}
Goldberg quotes an earlier paper by Freud \cite{PF} which is quite instructive in the
present context. Freud extends even earlier work by Weyl \cite{HW}
(Paragraph 33, p.217) and realizes that
\be\label{frd1}
2\sqrt{-g}U^i_k=\delta^i_k\sqrt{-g}(R^\mu_\mu+G)-2\sqrt{-g}R^i_k+
               (\Gamma^i_{\mu\nu}\partial_k \tilde{g}^{\mu\nu}
	       -\Gamma^\nu_{\mu\nu}\partial_k \tilde{g}^{\mu i})
\ee
can be written as a total divergence
\be\label{frd2}
2\sqrt{-g}U^i_k=\partial_\nu\lbrace 
            \delta ^i_k(\tilde{g}^{\mu\nu}\Gamma^\rho_{\mu\rho}
                       -\tilde{g}^{\mu\rho}\Gamma^\nu_{\mu\rho})
           +\delta^\nu_k(\tilde{g}^{\mu\rho}\Gamma^i_{\rho\mu}
                       -\tilde{g}^{i\rho}\Gamma^\mu_{\rho\mu})
		       -(\tilde{g}^{\mu\nu}\Gamma^i_{\mu k}
                       -\tilde{g}^{\mu i}\Gamma^\nu_{\mu k})\rbrace
\ee
This sum can be represented as determinant
\begin{align}\label{frd3}
2\sqrt{-g}U^i_k= 
\begin{array}{|ccc|}	
       \delta^i_k &\delta^n_k        &\delta^\mu_k\\
\sqrt{-g}g^{i\rho}&\sqrt{-g}g^{n\rho}&\sqrt{-g}g^{\mu\rho}\\
\Gamma^i_{\rho\mu}&\Gamma^n_{\rho\mu}&\Gamma^\mu_{\rho\mu},
\end{array} 
\end{align}
Equations (\ref{frd1}),(\ref{frd2}),(\ref{frd3}) which are just copied from 
Freud's paper require quite some explanations. First to notation: indices 
which are contracted (i.e.\ are to be summed over) are written in Greek 
letters. Indices which are not contracted in latin one's. (In order to 
reproduce (\ref{frd1}), (\ref{frd2}) from (\ref{frd3}) one puts $n=\nu$ 
and performs the contraction resulting from 
$\delta^\mu_k$.) Second in content. First we reorder (\ref{frd1}), then we 
replace $G$ by $\mathcal{L}$ and identify the last bracket in (\ref{frd1}) 
as Lagrangian derivative.
\begin{align}\label{frd4}
2\sqrt{-g}U^i_k=&2(\frac{1}{2}\delta^i_k\sqrt{-g}R^\mu_\mu-\sqrt{-g}R^i_k)\\
                &+\delta^i_k G+(\Gamma^i_{\mu\nu}\partial_k\tilde{g}^{\mu\nu}
		               -\Gamma^\nu_{\mu\nu}\partial_k\tilde{g}^{\mu i}\\
2\sqrt{-g}U^i_k=&-2(-\frac{1}{2}\delta^i_k\sqrt{-g}R^\mu_\mu+\sqrt{-g}R^i_k)\\
 &+\delta^i_k\mathcal{L}+\partial_k g^{\rho\sigma}
	   \frac{\partial\mathcal{L}}{\partial(\partial_i g^{\rho\sigma})}
\end{align}
This lhs is called ``superpotential'' for the simple reason that it
represents the energy-momentum density of the gravitational field
in terms of a Lagrangian and its equation of motion; i.e.\
the difference of the standard EH equation of motion term and the
canonical EMT (rather pEMT) belonging to the Lagrangian $\mathcal{L}=G$.\\
Exactly the same formula (one can equate term by term, once one has identified
the respective $t^\nu_\mu$), is the starting point for Trautman 
\cite{AT}.

\begin{align}\label{Trtn}
c_3\kappa^{-2}\sqrt{-g}(R_\mu^\nu-\frac{1}{2}g_\mu^\nu R)=& 
	\frac{1}{2}c_3\kappa^{-2}(\sqrt{-g}\,t_\mu^\nu
		 +\partial_\lambda A_\mu^{\phantom{\mu}\lambda\nu})\\
A_\mu^{\phantom{\mu}\lambda\nu}
\equiv&\sqrt{-g}g^{\sigma[\rho}\delta^\nu_\mu g^{\lambda]\tau}
                  g_{\rho\sigma,\tau}\\
t^\nu_\mu\equiv& \delta^\nu_\mu\sqrt{-g}\mathcal{L}'_{\rm EH}
	-\partial_\mu g^{\rho\sigma}
	\frac{\partial(\sqrt{-g}\mathcal{L}'_{\rm EH})}
	     {\partial(\partial_\nu g^{\rho\sigma})}\\
	\mathcal{L}'_{\rm EH}\equiv&g^{\mu\nu}
	(\Gamma^\rho_{\mu\nu}\Gamma^\lambda_{\rho\lambda}	     
	-\Gamma^\lambda_{\mu\rho}\Gamma^\rho_{\nu\lambda})	     
\end{align}

Trautman refers to Goldberg, where it is however not so easily identified.\\

Trautman continues the story. He formulates within general relativity
boundary conditions under
which gravitational radiation is absorbed and emitted and shows that the
superpotential term leads to a physical energy-momentum operator $P_\mu$ once 
one studies
the relevant integral between hypersurfaces. This integral is invariant with
respect to those linear transformations under which the superpotential term is
covariant. Choosing these hypersurfaces reduces the general covariance, but
in a way which is known via the specific hypersurfaces: different experimenters
can communicate these data. The remaining covariance is then just given by those
linear coordinate transformations under which the superpotential terms are 
covariant.
But these leave the integral invariant.
The mentioned integral is sufficient to provide physical meaning.
Hence, for sure, the densities of energy-momentum referred to in the pEMT 
have no physical significance per se, the flux through the hypersurfaces 
however has. Weyl \cite{HW}, p.\ 271 actually pointed out precisely this 
fact for the simplified version he has studied then.\\

\subsection{Combination of matter and gravity sector}
A first step to the combined system has been undertaken in \cite{EKKSI} and 
\cite{EKKSII}. There the study of conformal transformation properties of 
matter EMT's has been
performed via coupling of a matter EMT to an external, i.e.\ non-propagating
field $h^{\mu\nu}$ which then -- due to imposing corresponding contact terms
for this field -- turned indeed out to be the gravitational field.
To be concrete: the algebra (\ref{lgbrct}) has been enlarged by contributions
of the field $g^{\mu\nu}=h^{\mu\nu}+\eta^{\mu\nu}$. The respective WI 
operator then reads
\begin{align}\label{wdtp}
W(\varphi,g;a)\equiv& W(\varphi;a)+W(g^{\mu\nu};a)\\
	W(\varphi;a)\equiv \int a^\mu w^T_\mu\Gamma\equiv&\int a^\mu( \partial_\mu\varphi\frac{\delta\Gamma}{\delta\varphi}
	-\frac{1}{4}\partial_\mu(\varphi\frac{\delta\Gamma}{\delta\varphi}))\\
W(g;a)\equiv&\int( a^\rho\partial_\rho g^{\mu\nu}\frac{\delta}{\delta g^{\mu\nu}}
	-\partial_\lambda a^\mu g^{\lambda\nu}\frac{\delta}{\delta g^{\lambda\nu}}
	-\partial_\lambda a^\nu g^{\mu\lambda}\frac{\delta}{\delta g^{\mu\lambda}})
\end{align}	
Obviously these contact terms are just identical in form with the
$s$ transformation of $\varphi,g^{\mu\nu}$. \footnote{In \cite{EKKSIII} it 
has been shown that these contact terms (and the covariant version for 
$g_{\mu\nu}$) are uniquely determined by the algebra -- up to field 
redefinitions of $h^{\mu\nu}$ as a
function of itself.} Hence we covariantize the matter action
(\ref{mttct})  arriving at 
\be\label{cvm}
\Gamma({\rm matter})= 
\frac{c_k}{2} 
          \int(-g)^{1/4}g^{\mu\nu}\mathcal{D}_\mu\varphi\mathcal{D}_\nu\varphi
	 -\frac{\lambda}{4!}\int\varphi^4 
	 +\frac{c_{\rm R}}{2}\int (-g)^{1/4}\varphi^2R
\ee
The peculiar form of the $\lambda$ term as an $s$-invariant originates from
the ``unusual'' form of $s\varphi\,$; likewise $\mathcal{D}_\mu\varphi=
\partial_\mu\varphi-\frac{1}{8}\partial_\mu ln(-g)\varphi$. As an other invariant 
solution of the WI, which contains matter fields and is compatible with power 
counting four we added the ``non-minimal'' term with coefficient $c_{\rm R}$.\\ 
For constant $a^\rho$, ``rigid invariance'', we have
\be\label{wdtp2}
(W(g;a)+W(\varphi;a)\Gamma
	=a^\rho\int(\partial_\rho\varphi\frac{\delta\Gamma}{\delta\varphi}
	+\partial_\rho g^{\mu\nu}\frac{\delta\Gamma}{\delta g^{\mu\nu}})=0,
\ee	
for
\be\label{tgm}
		\Gamma =\kappa^{-2}\int\sqrt{-g}R+\Gamma({\rm matter})\\
\ee
The respective conserved current is the EMT for the combined system. We can
find it from the local WI
\begin{align}\label{ccs}
	(w_\mu(\varphi)+w_\mu(g))\Gamma	\equiv
   \partial_\mu\varphi\frac{\delta\Gamma}{\delta\varphi}
	-\frac{1}{4}\partial_\mu(\varphi\frac{\delta \Gamma}{\delta\varphi})\\
	+\partial_\mu g^{\mu'\nu'}\frac{\delta\Gamma}{\delta g^{\mu'\nu'}}
   +\partial_\lambda(g^{\lambda\nu'}\frac{\delta\Gamma}{\delta g^{\mu\nu'}}
+g^{\mu'\lambda}\frac{\delta\Gamma}{\delta g^{\mu'\mu}})=
	                                     \partial_\nu\Lambda^\nu_\mu\\
(w_\mu(\varphi)+w_\mu(g))\Gamma=-\partial_\nu T^\nu_\mu
    -2\partial_\lambda(\sqrt{-g}(R^\lambda_\mu-\frac{1}{2}\delta^\lambda_\mu R))\\
	(w_\mu(\varphi)+w_\mu(g))\Gamma=-\partial_\nu(T^\nu_\mu+I^\nu_\mu
	+2\sqrt{-g}(R^\lambda_\mu-\frac{1}{2}\delta^\lambda_\mu R))
\end{align}
Here we have firstly recovered as total derivative contact terms all 
contributions 
which occur in the $s$-variations and secondly admitted a total derivative
on the rhs, which is then specified.\\
Inserting (\ref{frd3}) we have
\be\label{ccs2}
(w_\mu(\varphi)+w_\mu(g))\Gamma
       =-\partial_\nu(T^\nu_\mu+I^\nu_\mu+t^\nu_\mu+\sqrt{-g}U^\nu_\mu)
\ee
The improvement term $I^\nu_\mu$ and the superpotential term 
$U^\nu_\mu$ are, of course, annihilated
by the partial derivative in front, but should indicate what we have to expect
as ambiguity for the currents themselves. On shell the lhs vanishes because it
consists of contact terms: the sum of the currents is strictly conserved.\\
To repeat in other terms: The  complete w's, i.e.\ the sum $w(\varphi)+w(g)$,
generate the divergence of the matter current $T_{\mu\nu}$ of \cite{EKKSI}
and in addition a total derivative term of the Einstein tensor. This in turn 
yields the divergence of the pure gravity current $t_{\mu\nu}$ (its canonical
version).

When going over from the divergences to the currents themselves we are 
confronted with the fact that they are not uniquely determined: we may add 
terms which vanish identically when applying $\partial_\nu$ -- the 
improvement, respectively superpotential terms. For the matter
current that is the covariantized version of (\ref{mprvt}). For the  gravity 
current this is the superpotential term either in form of (\ref{frd2}) or in 
form of (\ref{Trtn}) -- they are the same. In particular it becomes clear what
the total derivative terms in $w(g)$ were good for: they yielded the Einstein
tensor part which combines appropriately to the sum of gravity (pseudo)tensor
+ superpotential term .\\

In the following formula we present for completeness the matter EMT, as
given in \cite{EKKSI}. The origin
of the different terms can be read of from the coefficients: $c_k$ from
matter kinetic term; $c_{\rm R}$ from non-minimal term; $c$ from 
improvement term (put in by hand). 
\begin{align}\label{clsemt}
T_{\mu\nu}(\varphi,h)
	=&(-g)^{1/4}(c_k(\mathcal{D}_\mu\varphi\mathcal{D}_\nu \varphi
       -\frac{1}{2}g_{\mu\nu}g^{\rho\sigma}
	\mathcal{D}_\rho\varphi\mathcal{D}_\sigma\varphi\\
&-\frac{1}{4}g_{\mu\nu}g^{\rho\sigma}\varphi
	(\mathcal{D}_\rho\mathcal{D}_\sigma
	-\Gamma^\lambda_{\rho\sigma}\mathcal{D}_\lambda)\varphi)\\
&-2c(\mathcal{D}_\mu\varphi\mathcal{D}_\nu\varphi
	+\varphi(\mathcal{D}_\mu\mathcal{D}_\nu
	 -\Gamma^{\lambda}_{\mu\nu}\mathcal{D}_{\lambda})\varphi\\
&-g_{\mu\nu}g^{\rho\sigma}\mathcal{D}_\rho\varphi\mathcal{D}_\sigma\varphi
	 -g_{\mu\nu}g^{\rho\sigma}\varphi(\mathcal{D}_\rho\mathcal{D}_\sigma
	 -\Gamma^{\lambda}_{\rho\sigma}\mathcal{D}_{\lambda})\varphi)\\
	&+c_R(R_{\mu\nu}-\frac{1}{4}Rg_{\mu\nu})\varphi^2)		 
\end{align}

\section{Contributions of higher derivative terms}
The above section was devoted to the case of pure EH + matter. When going to 
higher orders of perturbation theory the quest for power counting 
renormalizability forces us to 
add higher derivative terms. Hence we have to study now, already in tree 
approximation, how those contribute to energy-momentum tensors, 
superpotentials and improvement.\\

Since closed expressions, say in terms of the Riemann tensor, require 
considerable technical work, we
expand in the number of $h^{\mu\nu}$-fields. The EH case teaches us that
superpotentials, which arise in the context of the equation of motion, come 
along with linear terms in $h^{\mu\nu}$, whereas contributions to EMT's start 
with bilinear terms. Thus absence or presence of superpotential terms can 
be deduced by expanding the action 
$\int\sqrt{-g}(c_2 R^2+c_1R^{\mu\nu}R_{\mu\nu})$ up to order
two in $h^{\mu\nu}$ and then by looking into the respective equation of 
motion, i.e.\ linear terms.\\
The relevant explicit expressions have in fact been worked out already in 
\cite{pottel2021perturbative},
eqs. (17), (18), in projector form 
(notation $P^{(r)}_{\rm KL}, r=0,2; K,L=T,P$, explicit form s. appendix
\cite{pottel2021perturbative}). Rewritten from momentum to configuration
space they read\\

 \begin{align}\label{bltrs}
	 \int&\sqrt{-g}(c_1R^{\mu\nu}R_{\mu\nu}+c_2 R^2)_{|{\rm bilin}(h)} 
       \sim \int h^{\mu\nu}\Box\Box\lbrace-c_1 P^{(2)}_{\rm TT}	 	 
           +(3c_2+c_1) P^{(0)}_{\rm TT}\rbrace_{\mu\nu\rho\sigma}h^{\rho\sigma}\\
        = \int& h^{\mu\nu}\lbrace -\frac{c_1}{2}\lbrack
	 (\eta_{\mu\rho}\Box-\partial_\mu\partial_\rho)	 
	 (\eta_{\nu\sigma}\Box-\partial_\nu\partial_\sigma)	 
	 +(\eta_{\mu\sigma}\Box-\partial_\mu\partial_\sigma)	 
	 (\eta_{\nu\rho}\Box-\partial_\nu\partial_\rho)\\
	 &\qquad -\frac{2}{3}(\eta_{\mu\nu}\Box-\partial_\mu\partial_\nu)	 
     (\eta_{\rho\sigma}\Box-\partial_\rho\partial_\sigma)\rbrack \\
	&\qquad +(3c_2+c_1)\frac{1}{3}(\eta_{\mu\nu}\Box-\partial_\mu\partial_\nu)
 (\eta_{\rho\sigma}\Box-\partial_\rho\partial_\sigma)\rbrace h^{\rho\sigma}
 \end{align}

(On the rhs an irrelevant overall numerical factor has been subpressed.) 
The transversality of these {\sl local} projectors is maintained when going 
over via
$\delta/(\delta h^{\mu\nu})$ to the equation of motion. Hence it is clear 
that these terms can not form a superpotential, since one can not extract 
an overall derivative in a local fashion. \\ 
Looking back into the EH situation one finds that there
the decomposition of the bilinear terms into projectors requires the spin zero
$r=0, K=L=T$ to cancel the spin 2 $r=2, K=L=T$ terms 
$\partial_\mu\partial_\nu\partial_\rho\partial_\sigma /\Box$ in order to 
maintain locality in the sum of all terms. Hence the transverse projector 
structure is destroyed and pulling out an overall derivative is possible.\\

A similar observation can be made for the matter contribution. The improvement
term is transverse, hence it is not a superpotential.\\

A technical remark is in order. When deriving equations of motion and 
the like one neglects as a rule total derivative terms. Hence to any chosen 
version of them, e.g.\ by defining a basis in terms of monomials, there 
exists a vast number
of total derivative terms which can mask a transverse structure. Hence 
picking out a specific one might lead to seemingly inconsistent versions. 
However, once one has found, say, a version in transverse projectors, then
the conclusions drawn from it are valid, since the multitude of total 
derivatives
forms equivalence classes. There not every member must have the same 
transversality property: this is not the property of a class, but of a 
representative of the class.
Similarly for superpotential. That means being a superpotential is not a class
property, but that of a specific member. And precisely that characterizes
the specifically relevant representative. (Here it is important that we work
in tree approximation; in higher orders one would need a characterization by
mathematical tools, like special conformal symmetry.)\\

If we follow the arguments of Trautman as far as radiation is concerned, 
the conclusion is quite interesting: whereas EH can, due to the 
superpotential, produce gravitational radiation, the higher derivative part 
of the complete action can not. In the context of EH + hds, which is the only
viable candidate for higher order perturbative extension, this result is 
highly welcome. 
From \cite{PS23} we know that beginning with one-loop the hds looses particle 
interpretation, hence the violation of unitarity is reduced to tree 
contributions.
But on tree level the above discussion just shows that the hds does not 
lead to radiation either. Hence the model EH + hds is a viable realization of
quantum gravity in a perturbative fashion. The higher derivatives are 
needed for mathematical consistency, but they do not ruin the physical 
interpretation.\\ 
In a technical sense, we observe that the harmlessness of hds in higher 
orders of perturbation theory \cite{PS23} is based on the $S$-matrix: the 
scattering operator. The physical irrelevance in tree approximation, which was
left open there, is derived here from looking at EMT: a different operator.\\

\subsection*{Acknowledgement}
The author is deeply indebted to Denson Hill for having provided him with his
survey quoted in the references. Without this the present paper could not
have been written.\\

\bibliographystyle{alpha}
\bibliographystyle{operator_weyl}

\begin{thebibliography}{BPS91}


\bibitem[PF]{PF}
Ph. Freud.
\newblock Über die Ausdrücke der Gesamtenergie und des Gesamtimpulses eines 
materiellen Systems in der Allgemeinen Relativitätstheorie.
\newblock {\em Annals of Mathematics}, 40: 417--419, 1939 

\bibitem[EKKSI]{EKKSI}
E. Kraus and K. Sibold.
\newblock Conformal transformation properties of the energy momentum tensor
in four dimensions.
\newblock {\em Nuclear Physics B}, 372: 113--144 , 1992  

\bibitem[EKKSII]{EKKSII}
E. Kraus and K. Sibold.
\newblock Local couplings, double insertions and the Weyl consistency condition.
\newblock {\em Nuclear Physics B}, 398: 125--154 , 1993  

\bibitem[EKKSIII]{EKKSIII}
E. Kraus and K. Sibold.
\newblock The general transformation law of the gravitational field via Noether's
procedure.
\newblock {\em Annals of Physics}, 219: 349--363 , 1992  


\bibitem[PS21]{pottel2021perturbative}
S.~Pottel and K.~Sibold.
\newblock Perturbative quantization of Einstein-Hilbert gravity embedded in a
  higher derivative model.
\newblock {\em Physical Review D}, 104(8):086012, 2021.

\bibitem[PS23]{PS23}
S.~Pottel and K.~Sibold.
\newblock Perturbative quantization of Einstein-Hilbert gravity embedded in a
  higher derivative model II.
\newblock arxiv:2308.15824v2 [hep-th]

\bibitem[KS24]{KS24}
K.~Sibold.
\newblock Einstein-Hilbert gravity, higher derivatives and a scalar matter field.
\newblock arxiv:2405.00528 [hep-th]

\bibitem[HN]{HN}
C. Denson Hill and Pawel Nurowski.
\newblock How the green light was given for gravitational wave search.
\newblock arxiv:1608.08673v1 [physics.hist-ph] 30 Aug 2016

\bibitem[AT]{AT}
Andrzej Trautman.
\newblock Radiation and boundary conditions in the theory of gravitation.
\newblock arxiv:1604,03145v1 [gr-qc] 11 Apr 2016

\bibitem[JG]{JG}
J. N. Goldberg.
\newblock Conservation laws in general relativity.
\newblock Phys.\ Rev.\ {\bf 111} 315--320 (1958).

\bibitem[HW]{HW}
Hermann Weyl.
\newblock Space, Time, Matter.
\newblock Dover Publications, New York 1959 (from 4th edition 1922) 

\end{thebibliography}










\end{document}